\documentclass[12pt]{article}
\usepackage{latexsym}
\usepackage{epsfig}

\newcommand{\del}{\partial}
\newcommand{\beq}{\begin{eqnarray}}
\newcommand{\eeq}{\end{eqnarray}}
\newcommand{\be}{\begin{eqnarray*}}
\newcommand{\ee}{\end{eqnarray*}}

\newcommand{\ra}{\rightarrow}

\begin{document}

\centerline{\Large\bf{Oskar Klein and the fifth dimension\footnote{\small{Contribution to the 250 years celebration of the Transactions of the Royal Norwegian Society of Sciences and Letters, no.4, 2011.}}}}
\bigskip
\centerline{ Finn Ravndal}
\bigskip
\centerline{\it Department of Physics, University of Oslo, 0316 Oslo, Norway.}

\begin{abstract}

\small{After a short biographical summary of the scientific life of Oskar Klein, a more detailed and hopefully didactic presentation of his derivation of the relativistic Klein-Gordon wave equation is given. It was a result coming out of his unification of electromagnetism and gravitation based on Einstein's general theory of relativity in a five-dimensional spacetime. This idea had previously been explored by Kaluza, but Klein made it more acceptable by suggesting that the extra dimension could be compactified  and therefore remain unobservable when it is small enough. }

\end{abstract}

\begin{figure}[htb]
  \begin{center}
    \epsfig{figure=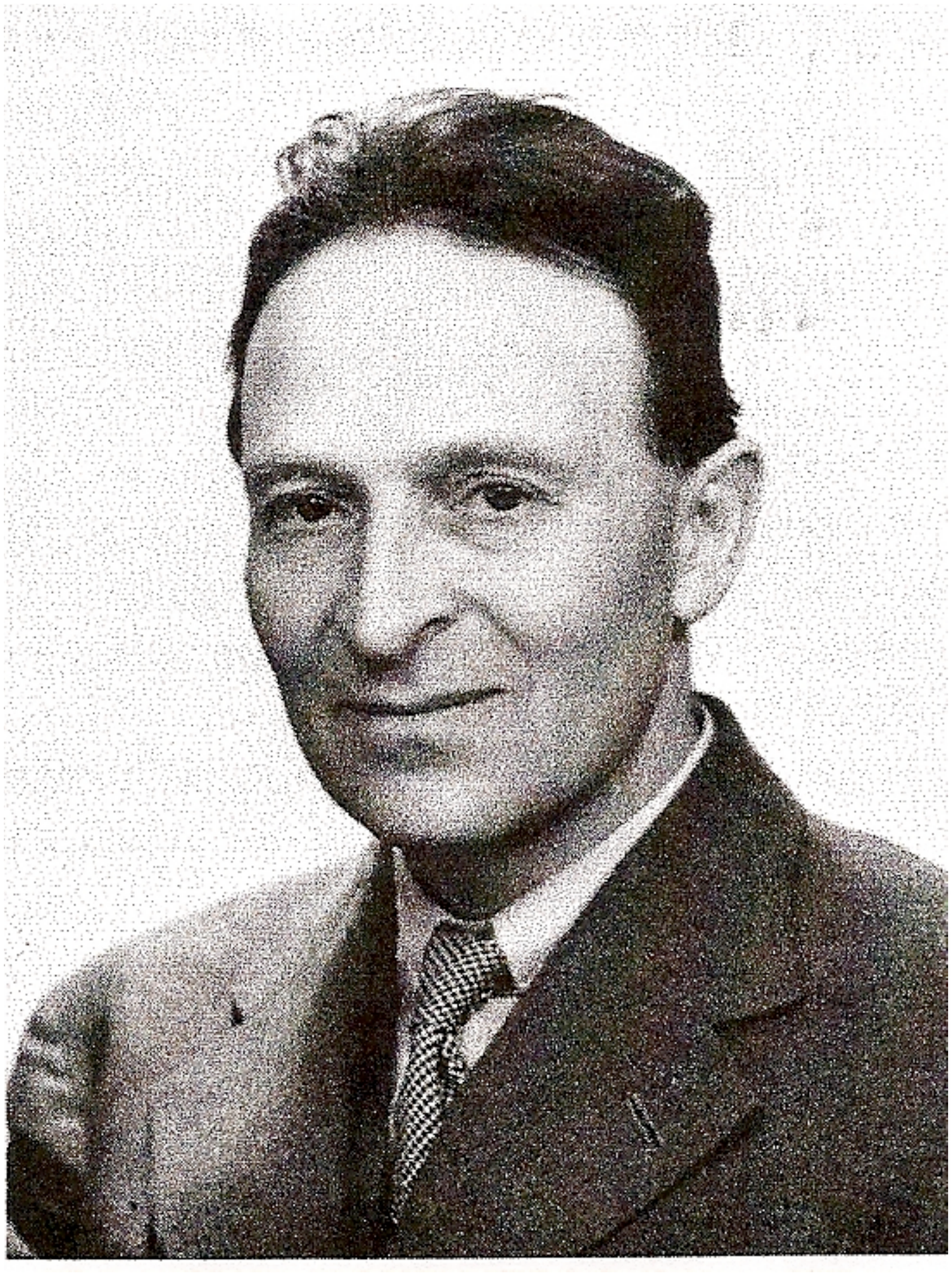,height=80mm}
  \end{center}
  \vspace{-4mm}            
\end{figure}
Besides the famous Danish physicist Niels Bohr, the founder of quantum mechanics, there are probably no other physicist in Scandinavia with his name connected to so many different, fundamental  physical theories and effects as the Swedish physicist Oskar Klein.  In chronological order we have the Klein-Gordon equation for relativistic quantum fields, the Kaluza-Klein theory for unification of electromagnetism and gravity, the Klein-Nishina formula for the Compton scattering cross-section, the Klein paradox in relativistic quantum mechanics, the Klein-Jordan quantization of many-particle systems and Klein's lemma for the quantum explanation of the second law of thermodynamics. Each of these contributions represents today major research fields which have seen great developments from the original ideas set forth by Klein. 

Oskar Klein was born in 1894 in Stockholm where he also got his early education. Already in high school he got in touch with the great chemist Arrhenius who was a friend of the family and continued working for him after he started his university studies. His first paper in 1917  was on the dielectric constants  of various solvents. The same year he met H. Kramers who came from Copenhagen to visit Arrhenius. Kramers worked with Bohr on atomic theory and was his first foreign collaborator. Klein took great interest in these new quantum ideas and with a fellowship he could himself  start to work with Bohr in 1918. In 1920 came S. Rosseland from Norway as the third, foreign collaborator to Bohr who then planned a new and bigger institute on the Blegdamsvej in Copenhagen. Already in the fall of the same year Klein and Rosseland had finished a joint paper on the opacities of stellar atmospheres. It is still well known and one of the very first papers where quantum ideas were used in astrophysics. In the period 1923 - 1925 he stayed at the University of Michigan in Ann Arbor and thus missed out on the emergence of modern quantum mechanics then taking place in Europe. But coming back, he started his most productive period where he explored several of the basic consequences of the new physics. Most of his important contributions mentioned above, were made in the short period 1926 - 1929. In 1930 he was appointed a full professor of theoretical physics at Stockholm University. In Norway he was a personal friend of the late H. Wergeland at the Institute for Theoretical Physics at NTH in Trondheim where he visited. In 1958 he published there in the proceedings of the Royal Academy a paper about  Dirac particles in a curved spacetime\cite{Ak}.  His last visit in Trondheim was in 1963. He died in 1977.

If one is trying to look for what probably occupied him at the deepest level and inspired him throughout his life, it is perhaps his quest to unify modern quantum principles with Einstein's theory of relativity. This endeavour led him in the first place to the extension of the Kaluza five-dimensional theory of gravity and electromagnetism and its generalizations as we know it today.  As a direct by-product this also led to the relativistic, quantum-mechanical wave equation which today carries his name. Later in his career the same ideas led him close to what we today call the unification of electroweak nuclear  forces in terms of massive non-abelian gauge fields\cite{Gross}  It is therefore perhaps appropriate that we in this expos\'e of his scientific contributions concentrate on this unification of quantum mechanics and general relativity.  And it was in particular quantum physics in curved spacetimes with one extra dimension that brought him the most lasting fame.

The first, relativistic theory of gravtitation was proposed in 1913 by the Finnish physicist Gunnar Nordstr\"om\cite{Nordstrom}. It was described by a scalar field as in Newton's non-relativistic theory. A year later he showed that this gravitational field and the ordinary electromagnetic fields could be combined in a unified field theory\cite{GN}. They were components of a new Maxwell vector field in a five-dimensional spacetime where the fifth component  contained the scalar field.

In 1919 the German mathematical physicist Theodor Kaluza considered Einstein's general theory of relativity in such a five-dimensional spacetime with coordinates $x^A = (x^\mu,y)$. Here are  $x^\mu$  the coordinates of our four-dimensional spacetime and $y\equiv  x^5$ is the new coordinate in the fifth dimension.  The gravitational field is the symmetric metric tensor $g_{AB}$ describing the geometric properties of the spacetime. Kaluza noticed that when this tensor field was independent of the new coordinate $y$, then the metric components $g_{5\mu}$ resulted in an interaction with the same properties as for an electromagnetic potential $A_\mu$ in our four-dimensional spacetime to lowest order. More accurately, using Einstein's theory he found that
\beq
                    g_{5\mu} = \kappa A_\mu         \label{em}
\eeq
where the constant  $\kappa^2 = 16\pi G/c^2$ when expressed in terms of Newton's gravitational constant $G$ and the velocity of light $c$. Einstein himself took much interest in this derivation and helped to get it published two years later\cite{Kaluza}. As with the previous proposal of Nordstr\"om, no one knew in what direction to look in order to observe this extra dimension.

When we introduce a set of basis vectors ${\bf e}_A$ in the five-dimensional spacetime, the metric components are given by the scalar products ${\bf e}_A\cdot{\bf e}_B = g_{AB}$. Our four-dimensional spacetime must be orthogonal to the basis vector ${\bf e}_5$ in the fifth dimension. It is therefore spanned by the four basis vectors
\beq
                {\bf e}_{\mu\perp} = {\bf e}_\mu -  {\bf e}_{\mu\parallel} = {\bf e}_\mu  - {{\bf e}_\mu\cdot{\bf e}_5\over{\bf e}_5\cdot{\bf e}_5}{\bf e}_5 =
                {\bf e}_\mu  -  {g_{\mu5}\over g_{55}}{\bf e}_5
\eeq
Klein\cite{Klein-1} assumed now that  $g_{55} = 1$ as also Kaluza had done. That corresponds to ignoring an addiditional, scalar field which is of no interest here\cite{Wehus-FR}. If our observed spacetime is assumed to be flat with the ordinary Minkowski metric $\eta_{\mu\nu}$, one has
\beq
             \eta_{\mu\nu} = {\bf e}_{\mu\perp}\cdot{\bf e}_{\nu\perp} = g_{\mu\nu} - {g_{5\mu}g_{5\nu}\over g_{55}} = g_{\mu\nu} - \kappa^2A_\mu A_\nu
\eeq
where we in the last step have introduced the electromagnetic field from (\ref{em}).

In this way we then have the following splitting of the five-dimensional metric
\beq
       {g}_{AB}= \left[\begin{array}{c}
                              \begin{tabular}{c | c}
                    $\eta_{\mu\nu}+ \kappa^2 A_{\mu}A_{\nu}$ & $\kappa A_{\mu}$ \\
                    \hline
                    $\kappa A_{\nu}$ & $1$ \\
                \end{tabular}
         \end{array}\right]                                                                    \label{covar}
\eeq
and correspondingly for the inverse matrix
\beq 
    {g}^{AB}= \left[\begin{array}{c}
                              \begin{tabular}{c | c}
                    $\eta^{\mu\nu}$ & $ -\kappa A^\mu$ \\
                    \hline
                    $ - \kappa A^\nu$ & $1 + \kappa^2 A_\mu A^\mu$ \\
                \end{tabular}
         \end{array}\right]           \label{contra}
\eeq
It is seen to satisfy ${g}_{AB} g^{BC} = \delta_B^C$ which are the components of the unit matrix. Here and in the following we use the Einstein convention which implies a summation over any index appearing twice in a mathematical expression.

The Lagrangian for a particle in the five-dimensional spacetime is
\beq
              L = {1\over 2}g_{AB}{\dot x}^A{\dot x}^B           \label{L}
\eeq
where the dot derivative is with respect to a parameter $\lambda$ describing the motion $x^A = x^A(\lambda)$. This motion will be a solution of the Euler-Lagrange equation
\beq
              {\del L\over \del x^A} - {d\over d\lambda}\left({\del L\over \del \dot{x}^A}\right) = 0   \label{EL}
\eeq
It is well known that the resulting equations describe a geodesic motion in this space. 

An alternative description of the particle can be obtained from the Hamiltonian function  $H = {\dot x}^A p_A - L$ where $p_A = \del L/\del{\dot x}^A $ is the canonical momentum for the particle. In our case it becomes $p_A = g_{AB}{\dot x}^B$.  With the assumption of Kaluza that the metric is independent of the fifth coordinate $y$, the fifth component of the momentum $p_5$ is now seen from (\ref{EL}) to be constant for the motion of the particle. It will in the following be convenient to write this constant as $p_5 = mc$ where $c$ is again the velocity of light.
 
 From the Lagrangian  (\ref{L}) now follows the Hamiltonian function
 \beq
               H = {1\over 2} g^{AB}p_Ap_B            \label{H}
 \eeq
It has  a constant value since the metric is independent of the parameter $\lambda$. In four-dimensional, relativistic mechanics this constant would correspond to minus the squared mass of the particle. Klein now assumed that the corresponding mass in five dimensions is zero, i.e. $H=0$. With the contravariant components of the metric from (\ref{contra}), the Hamiltonian (\ref{H}) then gives the equation
\beq
           (p_\mu - mc\kappa A_\mu)^2 + m^2c^2 = 0       \label{5-shell}
\eeq
which the particle momentum must satisfy. The massless particle in five dimensions thus appears as massive in four dimensions with mass $m= p_5/c$. In addition, it is seen to move in an electromagnetic potential $A_\mu$ to which it couples with charge $e = mc\kappa$.

Klein was inspired by  Schr\"odinger who the year before had proposed his quantum-mechanical wave equation\cite{Schr}. His derivation was also based on the Hamiltonian formalism in classical mechanics. In particular, Schr\"odinger  wrote the particle  momentum  as the gradient of the so-called principal function $S=S(x,y)$. It is essentially the action for the motion. In (\ref{H}) one can then set $p_A = \del S/\del x^A$ resulting in the differential Hamilton-Jacobi equation
\beq
              g^{AB}{\del S\over\del x^A} {\del S\over\del x^B} = 0        \label{HJ}
\eeq
When the implicit summation over the indices is done, it can be written on the same form as (\ref{5-shell}) with this substitution for the momenta. The classical motion is normal to the surfaces $S=const$ and therefore similar to the picture we have of light rays moving normal to their wave fronts in geometrical optics. This analogy was used by Schr\"odinger to propose that also in mechanics there should be a similar and more accurate wave description for which the Hamilton-Jacobi equation (\ref{HJ}) is just an approximation in a classical limit similar to what underlies geometrical optics. Needless to say, this proposed wave description of particle physics would soon be known as quantum mechanics.

For the relativistic motion in five dimensions Klein therefore proposed the invariant wave equation\cite{Klein-1}
\beq
             \del_A(g^{AB}\del_B)\Psi = 0     \label{waveq}
\eeq
where $\del_A = \del/\del x^A$. The wave function $\Psi = \Psi(x,y)$ can in general be a complex quantity for which the action function $S(x,y)$ is the phase of the wave. Thus we can write $\Psi = \exp(iS/\hbar)$ where the constant $\hbar$ has the same dimension as for mechanical action and turns out to be the Planck-Dirac constant. Inserting this wave function into (\ref{waveq}) and taking the classical limit $\hbar \ra 0$, one then recovers the Hamilton-Jacobi equation (\ref{HJ}). Since $p_5 = const$, we must have $S(x,y) = p_5y + S(x)$ and can thus write
\beq
            \Psi(x,y) = e^{ip_5y/\hbar}\Phi(x)           \label{wavef}
\eeq
where now $\Phi(x)$ should be Klein's wave function in four-dimensional spacetime. Writing again $p_5 = mc$, one then finds from (\ref{waveq}) that it must obey the differential equation
\beq
              \left[\left(\del_\mu - {i\over\hbar}eA_\mu\right)^2   + \left({mc\over\hbar}\right)^2 \right] \Phi(x) = 0    \label{KG}
\eeq
This is the  quantum wave equation used for  relativistic particles without spin and used today exactly on this original form. It is called the Klein-Gordon equation since W. Gordon\cite{Gordon} derived it the same year along very similar lines starting from the four-dimensional, classical equation (\ref{5-shell}).

It is Klein's idea of an extra dimension that distinguishes his approach from what Gordon and others did. One should naturally ask why he came upon this idea. He himself has said that he was attracted to the emergence of electromagnetism from such a geometrical point of view based on the Hamilton-Jacobi equation. He also acknowledges discussions with N. Bohr in this connection\cite{Klein-1}. It could be that they considered the new and mysterious quantum physics  in four dimensions as a reflection of simpler mechanics in higher dimensions.

A short time later Klein proposed that the extra dimension should be compact\cite{Klein-2}. In fact, it should be a circle with radius $R$. When it is smaller than what can be measured, one then no longer has the problem that it has not been seen. On a such a circle one comes back to the same point when the coordinate $y$ in (\ref{wavef}) is increased by $2\pi R$. Since the wave function then has the same value, one must have $p_5 = n\hbar/R$ where $n = 0, \pm 1, \pm 2, \cdots$. Klein found this result very intriguing since it would explain why all particle charges seem to be quantized in units of the electron's charge $e$. Calculating the fine structure constant $\alpha = e^2/4\pi\hbar c = 1/137$ with $e = mc\kappa$, one finds for $n=1$ that $\alpha = 4\hbar G/c^3R^2$ using the classical value $\kappa^2 = 16\pi G/c^2$. Thus $\alpha = 4(L_P/R)^2$ where $L_P = (\hbar G/c^3)^{1/2} = 1.62\times 10^{-33}\,$cm is the so-called Planck length. Thus $R = 23L_P$ and much too small even to be detected today. However, the  corresponding electron mass $m = \hbar/cR$ becomes $m = M_P/23$ where $M_P = (\hbar c/G)^{1/2} = 1.22\times 10^{19}\,\mbox{GeV}/c^2$ is the Planck mass. This is a factor $10^{22}$ larger than the physical mass of the electron!  Klein was for this reason a year later more pessimistic about his theory\cite{Klein-3}. However, today the situation is somewhat different. More advanced theories with one or more extra dimensions are today important in modern elementary physics and there is an  experimental program for their detection at the LHC accelerator at CERN.

\end{document}